# The Artscience of Planet Formation
# Import ArtScience.PlanetFormation as AATS

A protoplanetary disk is defined as a "rotating circumstellar disk of gas and dust surrounding a young newly formed star, a T Tauri star, or Herbig Ae/Be star". To an astrophysicist, each word in this definition carries meaning, implies specific physical properties, history and, most importantly, assumptions. To anyone outside the field of astrophysics the terms involved in the definition can be obscure. The assumptions which scientists had to adopt to build this otherwise precise description are difficult to unveil. Behind each assumption lies an intuition. An intuition which drove the scientists to explore the phenomenon in the first place, and later to characterize it using physics and math (more languages). The language, jargon, definitions, and assumptions are key, and sometimes unique, to a specific discipline. Communication between disciplines requires the translation of these languages, but most importantly, the sharing of underlying intuitions. This is an initial step to multidisciplinarity: the attempt to dialogue.

Seven years ago, I committed to a project called AATS, which stands for *art, astronomy, technology* and *society*[1]. The project was led by media artist Olaf Peña Pastene[2] in Santiago, Chile. Olaf approached the astronomers at a newly created research group, the Millennium ALMA Disk (MAD) nucleus, with the intention of building an alliance. I, an astrophysicist in the group, was particularly keen on exploring the art and science connection because I am also a musician. AATS was aimed at translating the new discoveries and *scientific knowledge* using contemporary/media art as a communication vehicle —this was the "art" in AATS. The "astronomy" was specific to protoplanetary disks: the place where we think planets are formed. There was a clear "technology" component: our group was in the first row to use the newly commissioned Atacama Large Millimeter/submillimeter Array (ALMA), the most powerful radio interferometer built thus far. The funding came from an outreach grant, which allowed for the transference of the AATS experience to the public – ergo, a "society" component.

Let us go back to protoplanetary disks for a second. "How did the Earth and the planets come to be?" is one of the oldest and at the same time one of the newest concerns of humanity. Before the 90s, there was no evidence of planets happening anywhere else in the Universe but in our Solar System. Along with a collection of asteroids, comets, moons, and dwarf planets, there were 8 planets orbiting the sun. Only in the early 90s, a planet which orbits a star that is not the sun (what we now call an "exoplanet") was discovered, changing a paradigm: there are planets outside the Solar System. Astronomers quickly invented new techniques and refined their methods in such a way that by min 2019 several thousand exoplanets are known.

The statistics are mind-blowing: every star in our Galaxy hosts, on average, at least one planet. Nature must be very efficient at making planets, yet we still lack crucial information to tell the story of how these planets are born. And this is why we study protoplanetary disks, to understand how planets form.

In 2013 we published a paper reporting new aspects of protoplanetary disk phenomena. The disk we were reporting on (HD142527) had a massive hole in the middle (Fig. 1, *left*). The cavity was com-pletely empty of solid particles (dust) but had copious amounts of gas in it. The gas was denser in two streamers that joined the outer parts of the disk with the vicinity of the central star. We interpreted these streamers as being channeled by two forming giant planets (Fig. 1, *right*). These new data gave us the possibility to model these young protoplanetary systems with hydrodynamics, the formalism that allows for the description of fluids. This discovery, and the approach of modelling it with hydrodynamics, fueled the first version of AATS in 2013.

AATS 2013 consisted of an immersive installation about the Origins of the Solar System (*Origen del Sistema Solar*). The experience involved entering an area under a dome (5m in diameter). Both unpublished ALMA observations and hydrodynamic simulations were projected onto the dome (see Fig. 2). The immersive area was surrounded by a quadraphonic sound system through which the data (observations and simulations) were being translated into sounds (sonification). The immersive experience was accompanied by several monitors showing a 10min video summarizing our knowledge of protoplanetary disks for the public.

AATS continued in 2014 as a day of "Art and Astronomy" which served as an initial framework for an incubator of artscience projects. Four projects were born, including data visualization experiments, two interactive installations exploring the discovery of exoplanets, and an exploration of ancient astronomical petroglyphs in the context of modern astrophysics via audiovisual improvisation. In 2017, Olaf Peña performed several concerts sharing the artscience material from AATS in an audiovisual format. AATS still exist today.

The artscience convergence in AATS was not straightforward. At first, it was difficult to find a common language. The media art usage of "scientific" terms and the astrophysical jargons probed to be ideal for *lost in translation* type situations. The words *data*, *observation*, and *theory*, amongst many other words, seem to have significantly different connotations. Dialogues between scientists and artists are no simple endeavors, even if the people involved have a mix of art and science backgrounds.

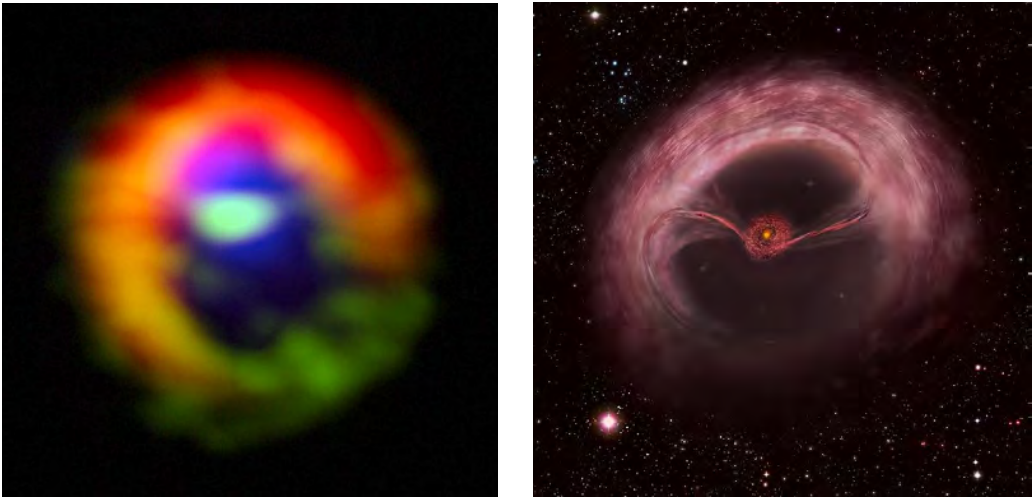

Figure 1 - *Left:* Observation of the protoplanetary disk HD142527 (Casassus et al. 2013, Nature, 493, 191). The star is near the center, the red and orange colors show cold dust particles and the green and blue show gas. *Right:* Artist impression of the data in the right, showing two protoplanets channeling streamers from the outer to the inner parts of the disk. Credit: Beatriz Buttazzoni.

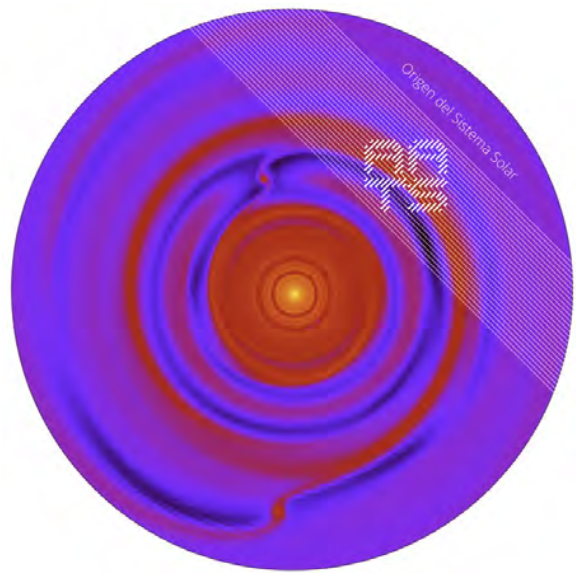

Figure 2 - Hydrodynamic simulation (numerical calculation) of how multiple protoplanets interact with the gas in a protoplanetary disk.

This realization was not self-evident: it took years of reflection and several artscience projects. In particular, recent collaborations with improvised movement[3] and music composition[4] have contributed greatly to understanding the complexity of artscience interactions. During the composition and recording of a Charango concerto based on the fundamental laws of the Universe, the composer Anya Yermakova and I agreed that, part of our artscience process must be building *shared* basic blocks of intuition about planet and stellar formation. In fact, the whole collaboration became about "fundamental laws" exactly for this reason, and the results can be seen at www.conciertocielos.cl/charango. Thus, with the current working hypothesis that the success of artscience interactions is about the process and about establishing a common intuitive ground of building blocks – further experiments await future versions of AATS.

*Acknowledgements:* The author is grateful to Anya Yermakova for insightful comments on the manuscript, Olaf Peña Pastene for the initial motivation to join AATS and Luiz Guilherme Vergara for the encouragement to write this work.

**Notas**

1 Visit www.sebaperez.io/aats

2 See some of his work here: https://vimeo.com/olaftone

3 Recreo Espacial project website: http://recreo.das.uchile.cl

4 Charango concerto website: www.conciertocielos.cl/charango